\documentstyle[preprint,aps]{revtex}
\begin{document}

\preprint{UTPT-95-26}

\title{Nonsymmetric Gravitational Theory as a String Theory}

\author{J. W. Moffat}

\address{Department of Physics, University of Toronto,
Toronto, Ontario, Canada M5S 1A7}

\date{\today}

\maketitle

\begin{abstract}
It is shown that the new version of nonsymmetric gravitational theory (NGT)
corresponds in the linear approximation to linear Einstein gravity theory and
antisymmetric tensor potential field equations with a non-conserved string
source current. The Hamiltonian for the antisymmetric field equations is
bounded
from below and describes the exchange of a spin $1^+$ massive vector boson
between open strings. The non-Riemannian geometrical theory is
formulated in terms of a nonsymmetric fundamental tensor $g_{\mu\nu}$. The weak
field limit, $g_{[\mu\nu]}\rightarrow 0$, associated with large distance
scales,
corresponds to the limit to a confinement region at low energies described by
an
effective Yukawa potential at galactic distance scales. The limit to this
low-energy
confinement region is expected to be singular and non-perturbative.
The NGT string theory predicts that there are no black hole event horizons
associated with infinite red shift null surfaces.
\end{abstract}

\pacs{ }

\narrowtext

\section{Introduction}

A new version of nonsymmetric gravitational theory (NGT) has recently been
published\cite{Moffat1,Moffat2,Moffat3,LegareMoffat1,Clayton1}, which was
shown to have a linear approximation free of ghost poles and tachyons with a
Hamiltonian bounded from below. The expansion to linear order in $g_{[\mu\nu]}$
about a fixed GR background also has a ghost-free Lagrangian with physical
asymptotic behavior. The theory produces good fits to
galaxy rotation curves and can explain gravitational lensing and cluster
dynamics
without appreciable amounts of dark matter\cite{MoffatSokolov1}.

In the following, we shall show that the linear approximation for weak fields
corresponds to the spin $2^+$ graviton linear equations of general relativity
(GR)
and to massive spin $1^+$ field equations. The source for the graviton field is
the
standard point particle energy-momentum tensor of GR, while the source for the
antisymmetric tensor field is a string source current for open strings which is
not
conserved.  The rigorous nonlinear action of NGT is
a nontrivial non-Riemannian geometrical unification of GR and string theory,
which
has as one of its predictions that black hole event horizons are not exected to
form
during gravitational collapse\cite{Moffat4,MoffatSokolov2}. The predictions
of the new NGT at cosmological scales is expected to produce a novel
dynamical scenario, as an alternative to the standard inflationary
model\cite{Moffat5}.

Clayton has developed a canonical Hamiltonian formalism for NGT,
and shown that the rigorous theory possesses six degrees of
freedom\cite{Clayton2}. He also
showed that the limit to the weak field linear theory for
$g_{[\mu\nu]}\rightarrow 0$
may be singular, i.e., a Lagrange multiplier associated with the skew fields
behaves
as $\sim 1/g_{[0i]}$ as $g_{[0i]}\rightarrow 0$ ($i=1,2,3$), so that the very
low-energy limit of NGT, going from six degrees of freedom of the rigorous
theory
to the three degrees of freedom of the linear theory, may be a singular limit.
In the following,  this limit is interpreted physically to be the weak field
galactic
scale limit of NGT,
for which the Newtonian and GR predictions fail to be valid. We shall argue
that
this is a low-energy string confinement limit for large distance scales of NGT,
which
is a {\it non-perturbative sector} of the theory in which three degrees of
freedom are
not excited.

\section{NGT Action and String Theory}

The nonsymmetric gravitational theory (NGT) is based on
the decomposition of the fundamental tensor $g_{\mu\nu}$:
\begin{equation}
\label{gequation}
g_{\mu\nu}=g_{(\mu\nu)}+g_{[\mu\nu]},
\end{equation}
where
\[
g_{(\mu\nu)}={1\over 2}(g_{\mu\nu}+g_{\nu\mu}),\quad g_{[\mu\nu]}=
{1\over 2}(g_{\mu\nu}-g_{\nu\mu}).
\]
The connection $\Gamma^\lambda_{[\mu\nu]}$ also has the decomposition:
\[
\Gamma^\lambda_{\mu\nu}=\Gamma^\lambda_{(\mu\nu)}
+\Gamma^\lambda_{[\mu\nu]}.
\]
The Lagrangian density takes the form:
\begin{equation}
\label{NGTaction}
{\cal L}_{NGT}={\cal L}_R+{\cal L}_M,
\end{equation}
where
\begin{equation}
{\cal L}_R={\bf g}^{\mu\nu}R_{\mu\nu}(W)-2\Lambda\sqrt{-g}
-{1\over 4}\mu^2{\bf g}^{\mu\nu}g_{[\nu\mu]}-{1\over 6}{\bf g}^{\mu\nu}
W_\mu W_\nu,
\end{equation}
where $g=\hbox{Det}(g_{\mu\nu})$, $\Lambda$ is a cosmological constant and
$\mu$
is a ``mass" associated with $g_{[\mu\nu]}$.
Moreover, ${\cal L}_M$ is the matter Lagrangian density ($G=c=1$):
\begin{equation}
{\cal L}_M=-8\pi g^{\mu\nu}{\bf T}_{\mu\nu}.
\end{equation}
Here, ${\bf g}^{\mu\nu}=\sqrt{-g}g^{\mu\nu}$ and $R_{\mu\nu}(W)$ is the
NGT contracted curvature tensor:
\begin{equation}
R_{\mu\nu}(W)=W^\beta_{\mu\nu,\beta} - {1\over
2}(W^\beta_{\mu\beta,\nu}+W^\beta_{\nu\beta,\mu}) -
W^\beta_{\alpha\nu}W^\alpha_{\mu\beta} +
W^\beta_{\alpha\beta}W^\alpha_{\mu\nu},
\end{equation}
defined in terms of the unconstrained nonsymmetric connection:
\begin{equation}
\label{Wequation}
W^\lambda_{\mu\nu}=\Gamma^\lambda_{\mu\nu}-{2\over 3}\delta^\lambda_\mu
W_\nu,
\end{equation}
where
\[
W_\mu={1\over 2}(W^\lambda_{\mu\lambda}-W^\lambda_{\lambda\mu}).
\]
Eq.(\ref{Wequation}) leads to the result:
\[
\Gamma_\mu=\Gamma^\lambda_{[\mu\lambda]}=0.
\]

We shall assume that $\Lambda=0$ and expand $g_{\mu\nu}$ about Minkowski
spacetime:
\[
g_{\mu\nu}=\eta_{\mu\nu}+{}^{(1)}h_{\mu\nu}+...,
\]
where $\eta_{\mu\nu}$ is the Minkowski metric tensor: $\eta_{\mu\nu}=
\hbox{diag}(-1, -1, -1, +1)$. We also expand $\Gamma^\lambda_{\mu\nu}$ and
$W^\lambda_{\mu\nu}$ in a similar manner, and we adopt the notation:
$\psi_{\mu\nu}={}^{(1)}h_{[\mu\nu]}$. Then, to first order of
approximation, we find that\cite{Moffat1,Moffat2}:
\[
\psi_\mu=-\frac{1}{2}{}^{(1)}W_\mu=\frac{16\pi}{\mu^2}{T_{[\mu\nu]}}^{,\nu},
\]
where
\[
\psi_\mu={\psi_{\mu\beta}}^{,\beta}=\eta^{\beta\sigma}\psi_{\mu\beta,\sigma}.
\]

The symmetric and antisymmetric field equations decouple to lowest order; the
symmetric equations are the usual Einstein field equations in the linear
approximation, while the skew equations are given by
\begin{equation}
\label{divFeq}
{F_{\mu\nu\lambda}}^{,\lambda}+\mu^2\psi_{\mu\nu}=16\pi T_{[\mu\nu]},
\end{equation}
where
\[
F_{\mu\nu\lambda}=\psi_{\mu\nu,\lambda}+\psi_{\nu\lambda,\mu}
+\psi_{\lambda\mu,\nu}.
\]
Eq.(\ref{divFeq}) can be written as
\begin{equation}
\label{boxequation}
(\Box+\mu^2)\psi_{\mu\nu}=16\pi(T_{[\mu\nu]}+{2\over
\mu^2}{T_{[[\mu\sigma],\nu]}}^{,\sigma}).
\end{equation}

The action has the form:
\begin{equation}
\label{Faction}
S=\int
d^4x\biggl(\frac{1}{12}F_{\mu\nu\lambda}F^{\mu\nu\lambda}-\frac{1}{4}\mu^2
\psi_{\mu\nu}\psi^{\mu\nu}+8\pi\psi^{\mu\nu}T_{[\mu\nu]}\biggr).
\end{equation}

The form of the field equation, Eq.(\ref{boxequation}), is the same as that
derived
by Kalb and Ramond from a string action\cite{KalbRamond}:
\begin{equation}
I=-\Sigma_a\mu_a^2\int(-d\sigma_a\cdot d\sigma_a)^{1/2}
+\Sigma_{a,b}g_ag_b\int d\sigma_a^{\mu\nu}d\sigma_{b\mu\nu}\Delta(s_{ab}^2),
\end{equation}
where $\Delta(s_{ab}^2)$ is a Green's function describing time-symmetric
interactions, and
\[
s_{ab}^2=(x_a-x_b)\cdot(x_a-x_b).
\]
Moreover,
\[
d\sigma_a^{\mu\nu}=d\tau_ad\xi_a\sigma_a^{\mu\nu},
\]
where
\[
\sigma^{\mu\nu}_a=\dot{x}_a^\mu
x_a^{\prime\nu}-x_a^{\prime\mu}\dot{x}_a^\nu
\]
and
\[
\dot{x}_a^\mu=\frac{\partial x_a^\mu}{\partial\tau_a},\quad x_a^{\prime\mu}=
\frac{\partial x_a^\mu}{\partial \xi_a}.
\]
The coupling constants $g_a$ have the units of mass,  $\mu_a$ is chosen to
make the action dimensionless in natural units and the sums are over all
strings. A string is a one-dimensionally extended object,
$x^\mu_a(\tau_a,\xi_a)$,
which is traced out by a world sheet in spacetime by the invariant parameters
$\tau_a$ and $\xi_a$. The action is manifestly parametrization invariant.

The current density $T_{a[\mu\nu]}$ has the form:
\begin{equation}
\label{Jcurrent}
T_{a[\mu\nu]}=g_a\int d\sigma_{a\mu\nu}\delta^{(4)}(y-x_a(\tau,\xi)).
\end{equation}
For the open string interactions the current is not conserved:
\begin{equation}
%% FOLLOWING LINE CANNOT BE BROKEN BEFORE 80 CHAR
{T^{[\mu\nu]}}_{a,\nu}(y)=g_a\int^{\tau_f}_{\tau_i}d\tau[\dot{x}^\nu_a(\tau,\xi)
\delta^{(4)}(y-x_a(\tau,\xi))]^{\xi=l}_{\xi=0},
\end{equation}
where $l$ is a constant with dimensions of a length. The non-conservation
of the source is due to the dependence of the right-hand side on the end points
of the
string.

The Hamiltonian obtained from the action, Eq.(\ref{Faction}), is bounded from
below for reasons similar to those that apply to the point particle action of
the
massive Maxwell-Proca theory.

A more general string action in spacetime can be written\cite{Green,Moffat6}:
\begin{equation}
I=-\frac{1}{4\pi\alpha^\prime}\int d\xi d\tau g_{\mu\nu}\sqrt{-g}g^{ab}
\partial_a X^\mu\partial_b X^\nu,
\end{equation}
where $g_{\mu\nu}$ is the nonsymmetric fundamental tensor,
$g=\hbox{Det}(g_{ab})\, (a,b=1,2)$ and $\alpha^\prime$ is related to the string
tension $T$ by $T=(2\pi\alpha^\prime)^{-1}$.

\section{The Low-energy Confinement Sector of the NGT String Theory}

The vibrations of the strings in the theory generate modes of excitation
corresponding to different field degrees of freedom. Clayton has
used a canonical Hamiltonian formulation of NGT to demonstrate that the
full non-linear theory possesses six degrees of freedom\cite{Clayton2}. The
diffeomorphism invariance of the action, Eq.(\ref{NGTaction}), reduces the
number of degrees of freedom of $g_{\mu\nu}$ from 16 to 12; their is no further
reduction of the degrees of freedom in the rigorous theory, owing to a lack of
further gauge invariance constraints in the antisymmetric sector. However, the
linear
approximation for weak fields is characterized by field equations with only 3
degrees of freedom, since the three $g_{[0i]}$ components can be gauged
away due to the gauge invariance of the kinetic energy term:
$\frac{1}{12}F_{\mu\nu\lambda}F^{\mu\nu\lambda}$ in the action,
Eq.(\ref{Faction}), with $\delta\psi_{\mu\nu}=\xi_{\mu,\nu}-\xi_{\nu,\mu}$. The
Lagrange multiplier associated with $\Gamma^\lambda_{[\mu\nu]}$, determined by
the $g$ and $\Gamma$ compatibility equation in the Hamiltonian
formulation, has a singular limit as $g_{[0i]}\rightarrow 0$, due to a
transition
from a derivative coupled kinetic energy in the rigorous theory to the gauge
invariant
kinetic energy of the linear approximation\cite{Clayton2}.

This feature of NGT can be understood by viewing the vibrations of the strings
as generating different energy scales. A basic scale is set by
$r_0=\mu^{-1}$; it is interpreted as the onset of the ``confinement"
distance scale. It corresponds to the low-energy scale for very weak
$g_{[\mu\nu]}$
fields, and is taken to be of galactic dimensions, $r_0\sim 25$ kpc. The
transition
from the non-confining to the confining energy region is characterized by a
reduction in the number of degrees of freedom in the ``effective" NGT field
theory
description of the string theory. Thus, an increase in energy as the strings
vibrate, excites the additional three degrees of freedom associated with
$g_{[0i]}$,
although these degrees of freedom are only measurable ``locally" in the
spacetime
structure, due to the short-range nature of the antisymmetric field
$g_{[\mu\nu]}$.
The singular limit of the theory, as $g_{[\mu\nu]}\rightarrow 0$, is due to the
non-perturbative nature of the confinement region.

An effective model of the the low-energy confinement limit has been
derived from the weak field point particle limit of NGT
\cite{MoffatSokolov1,LegareMoffat2}. The total radial acceleration on a test
particle in the weak field limit has the form:
\begin{equation}
\label{finalacceleration}
a(r)=-\frac{G_0M}{r^2}\biggl\{1+\sqrt{\frac{M_0}{M}}[1-\exp(-r/r_0)
(1+r/r_0)]\biggr\},
\end{equation}
where $M_0$ is a constant mass parameter, and the gravitational constant at
infinity
$G_{\infty}$ is defined by
\begin{equation}
\label{renormgrav}
G_{\infty}=G_0\biggl(1+\sqrt{\frac{M_0}{M}}\biggr),
\end{equation}
and $G_0$ is the Newtonian gravitational constant. Thus, the gravitational
constant
$G$ ``runs" with energy in analogy with the coupling constant
$\alpha_{\hbox{qcd}}$ in quantum chromodynamics.  In the high-energy
limit $M\rightarrow\infty$, we have $G_{\infty}\rightarrow G_0$.
The effective potential has the form for a range of $ r \leq r_0$:
\[
V(r)\sim \frac{a}{r}+b\ln r,
\]
where $a$ and $b$ are constants. This describes a phenomenological $1/r$ plus
a confining string potential.
The non-additive nature of the Yukawa contribution in
(\ref{finalacceleration}),
caused by the $\sqrt{M}$ factor, is related to the non-perturbative nature of
the
confinement limit.

Eq.(\ref{finalacceleration}) has been applied to galaxy
dynamics and good fits were obtained for small and large spiral galactic
rotation
curves, without assuming appreciable amounts of dark
matter\cite{MoffatSokolov1}.
It is also possible to explain gravitational lensing effects
and cluster dynamics without significant amounts of dark matter.

At somewhat higher energies -- corresponding to the scale of the solar system
--
the Newtonian law of gravity and the corrections due to GR are valid,
and the standard tests of Newtonian and GR theories will be predicted,
in agreement with observations. The weak equivalence principle is retained in
the new version of NGT, although the strong equivalence principle, which states
that the non-gravitational laws of physics will not be the same in different
local
frames of reference, will not be valid.

The string sources in NGT may also be responsible for the elimination of
black hole event horizons in gravitational collapse
\cite{Moffat4,MoffatSokolov2,CornishMoffat1,CornishMoffat2}. For strong fields
near the Schwarzschild radius, $r=2GM/c^2$, the static spherically symmetric
vacuum solution does not possess any null surfaces in the range
$0<r\leq\infty$,
which results in a coordinate invariant, finite red shift for collapsed
astrophysical
objects. Nonetheless, their can exist a collapsed, massive compact object
with a large but finite red shift that
simulates the putative observational evidence for black holes. Near the
Schwarzschild event horizon, the string theory is described by a high energy
or short distance scale, and the effects of the $g_{[\mu\nu]}$ fields become of
critical importance. The elimination
of black hole event horizons would remove the potential information loss
problem
at the classical level. The possibility
that string theory may eliminate black hole event horizons has also been
suggested
by Cornish\cite{Cornish}.

\section{Conclusions}

We have shown that NGT has to be associated with string theory, because the
source of the antisymmetric field $g_{[\mu\nu]}$ is naturally described by
strings
as demonstrated by Kalb and Ramond\cite{KalbRamond}.  This means
that NGT can be interpreted as a geometrical unification of Einstein gravity
and
bosonic string theory.
The weak field linear approximation corresponds to a non-perturbative limit
reached at galactic scales, where Newtonian and Einstein gravity fail to be
valid.
The transition from a Newtonian and Einstein potential energy to a
confinement potential energy, described by an effective point-like Yukawa
potential, predicts radically different gravitational dynamics at the galactic
scale,
exhibited observationally by the flat rotational velocity curves of spiral
galaxies. The
gravitational constant runs with energy in analogy to the
coupling constant in quantum chromodynamics, and the gravitational confinement
region at low energy is non-perturbative.

Another interesting consequence of NGT is at the cosmological scale, when the
$g_{[\mu\nu]}$
field can be a natural source of inhomogeneities in the early universe.  The
NGT field equations could dynamically evolve towards a solution close to the
standard big bang model with a small effective cosmological constant at the
present epoch, produced by the antisymmetric field.  Thus, the standard big
bang
model with a cosmological constant would act as an attractor for the solutions
of the
NGT field equations.

\acknowledgements

This work was supported by the Natural Sciences and Engineering Research
Council of Canada. I thank I. Yu. Sokolov for several helpful suggestions and
comments. I also thank M. A. Clayton, L. Demopoulos and P. Savaria for
helpful and stimulating discussions.

\end{document}